\documentclass[sigconf,screen]{acmart}

\usepackage{multirow}
\usepackage{tabularx}
\usepackage{geometry}
\usepackage{graphicx}     
\usepackage[T1]{fontenc}  
\usepackage{txfonts}
\usepackage{mathptmx}
\usepackage{hyperref}
\usepackage{xcolor}
\usepackage{textcomp}
\usepackage{booktabs}
\usepackage{enumitem}

\AtBeginDocument{%
  \providecommand\BibTeX{{%
    \normalfont B\kern-0.5em{\scshape i\kern-0.25em b}\kern-0.8em\TeX}}}

\setcopyright{acmlicensed}
\copyrightyear{2024}
\acmYear{2024}
\acmDOI{XXXXXXX.XXXXXXX}

\acmConference[ICER 2024]{2024 ACM Conference on International Computing Education Research}{August 13--15, 2024}{Melbourne, Australia}
  
\usepackage{listings}
\begin{document}

\title[Evaluating LLMs for tasks performed by Undergraduate Computer Science Students]{"Which LLM should I use?": Evaluating LLMs for tasks performed by Undergraduate Computer Science Students}






  
\author{Vibhor Agarwal}
\email{vibhor20349@iiitd.ac.in}
\affiliation{%
  \institution{IIIT Delhi}
  \city{New Delhi}
  \country{India}}

  
\author{Madhav Krishan Garg}
\email{madhav21333@iiitd.ac.in}
\affiliation{%
  \institution{IIIT Delhi}
  \city{New Delhi}
  \country{India}}

\author{Sahiti Dharmavaram}
\email{sahiti.dharmavaram2021@vitstudent.ac.in}
\affiliation{%
  \institution{VIT Vellore}
  \city{Tamil Nadu}
  \country{India}}


\author{Dhruv Kumar}
\email{dhruv.kumar@iiitd.ac.in}
\affiliation{%
  \institution{IIIT Delhi}
  \city{New Delhi}
  \country{India}}

\renewcommand{\shortauthors}{Anonymous et al}

\begin{abstract}
This study evaluates the effectiveness of various large language models (LLMs) in performing tasks common among undergraduate computer science students. Although a number of research studies in the computing education community have explored the possibility of using LLMs for a variety of tasks, there is a lack of comprehensive research comparing different LLMs and evaluating which LLMs are most effective for different tasks. 
Our research systematically assesses some of the publicly available LLMs such as Google Bard, ChatGPT(3.5), GitHub Copilot Chat, and Microsoft Copilot across diverse tasks commonly encountered by undergraduate computer science students in India. These tasks include code explanation and documentation, solving class assignments, technical interview preparation, learning new concepts and frameworks, and email writing. Evaluation for these tasks was carried out by pre-final year and final year undergraduate computer science students and provides insights into the models' strengths and limitations. This study aims to guide students as well as instructors in selecting suitable LLMs for any specific task and offers valuable insights on how LLMs can be used constructively by students and instructors.

\end{abstract}



\maketitle

\section{INTRODUCTION}
The emergence of ChatGPT in November 2022 by OpenAI garnered significant attention, manifesting in a record-breaking userbase expansion that surpassed 100 million active users by January 2023, as reported in \cite{chatgpt_record}. Developed as a large language model (LLM), ChatGPT was trained on an extensive dataset comprising internet sources, including but not limited to Wikipedia, research papers, articles, publicly accessible codebases on GitHub, and StackOverflow. Its capabilities extend to generating human-like responses for diverse prompts, ranging from code generation to concept elucidation and the creation of written works such as essays and emails.

The success of ChatGPT prompted the advent of analogous language models by various entities. Noteworthy among these are Google Bard, announced by Google in February 2023 and integrated with Google search and related services \cite{Bard_announcement}, Microsoft Copilot, unveiled by Microsoft in the same month, amalgamating Bing Search with GPT-4 models developed in collaboration with OpenAI \cite{Mehdi_2023}, and Github Copilot Chat, introduced by GitHub in March 2023, specifically tailored for coding-related tasks and built upon the GPT-4 architecture in collaboration with OpenAI \cite{Copilot_GPT_Announcement}. Open-source counterparts such as Meta Llama-2, unveiled in July 2023 \cite{llama-2_announcement}, have also joined this landscape.

Within the computing education research community, a number of research studies \cite{Denny_Kumar_Giacaman_2023,Finnie-Ansley_Denny_Becker_Luxton-Reilly_Prather_2022,Reeves_Sarsa_Prather_Denny_Becker_Hellas_Kimmel_Powell_Leinonen_2023, Šavelka_Agarwal_Bogart_Song_Sakr_2023, Wermelinger_2023, Finnie-Ansley_Denny_Luxton-Reilly_Santos_Prather_Becker_2023, Cipriano_Alves_2023, Ouh_Gan_Gan_Wlodkowski_2023, software_change, Malinka_Peresíni_Firc_Hujňák_Januš_2023} have delved into the potential utility of Large Language Models (LLMs) in undergraduate Computer Science courses. While acknowledging their prospective benefits, they also highlight concerns such as plagiarism, dissemination of incorrect knowledge, and the potential stifling of creativity if over-relied upon. Nonetheless, there is a general consensus that LLMs are useful in a wide variety of contexts. For regions like India where English is the language of instruction but not the native language, these LLMs can be even more useful for students who have limited proficiency in the English language \cite{joshi2023interviews, Budhiraja_Joshi_Challa_Akolekar_Kumar_2024}.

Yet, a dearth of comprehensive research persists regarding the comparative effectiveness of various LLMs for common tasks undertaken by undergraduate computer science students.
This paper addresses this gap by undertaking a quantitative and qualitative analysis of four publicly available LLMs such as Google Bard, ChatGPT(3.5), GitHub Copilot Chat, and Microsoft Copilot (integrated with Bing) across a wide spectrum of tasks, including code explanation and documentation, algorithmic problem-solving for interview preparation from platforms like Leetcode \cite{LeetCode}, assistance with coursework assignments in core Computer Science subjects such as Operating Systems, and Algorithm Design and Analysis, as well as in Social Sciences and Humanities subjects\footnote{Each undergraduate computer science student is required to take some Social Sciences and Humanities courses also as part of degree curriculum in many Indian universities.}, assistance in learning new concepts and frameworks and writing emails. Evaluation for these tasks was carried out by pre-final year and final year undergraduate computer science students and provides insights into each LLM's strengths and limitations.

Specifically, the paper addresses the following research question: \textbf{What are the strengths and weaknesses of ChatGPT(3.5)\footnote{We are using free version of ChatGPT that comes with GPT-3.5}, Microsoft Copilot\footnote{We are using balanced mode in Microsoft Copilot}, GitHub Copilot Chat, and Google Bard when addressing queries related to undergraduate computer science courses, encompassing both technical and non-technical aspects of the undergraduate curriculum?}
To the best of our knowledge, this paper is the first attempt at comprehensively evaluating multiple LLMs specifically in the context of tasks commonly performed by undergraduate computer science students in India.
\section{RELATED WORK}
Studies have shown that LLM \cite{Floridi_Chiriatti_2020} tools can offer diverse benefits to undergraduate computer science students. Early research by Becker et al. \cite{Becker_Denny_Finnie-Ansley_Luxton-Reilly_Prather_Santos_2023} explored a variety of aspects related to using AI code creation tools, such as Amazon CodeWhisperer \cite{AI}, DeepMind AlphaCode \cite{AlphaCode}, and OpenAI Codex \cite{OpenAICodex}. In their study Becker et al. explore the potential of LLM tools in various applications. Nevertheless, it also highlights the necessity to tackle ethical, bias, and security concerns for these tools to gain broader acceptance. Similar challenges and opportunities have also been explored in \cite{software_change,Denny_Kumar_Giacaman_2023,Malinka_Peresíni_Firc_Hujňák_Januš_2023, joshi2023chatgpt}.

Numerous research studies have been conducted to assess the accuracy of Large Language Models (LLMs) like OpenAI Codex \cite{OpenAICodex}, GPT-3 \cite{GPT3}, and ChatGPT (including versions GPT-3.5 and GPT-4) in producing solutions for programming tasks across various computer science courses. These include CS1 \cite{Denny_Kumar_Giacaman_2023,Finnie-Ansley_Denny_Becker_Luxton-Reilly_Prather_2022,Reeves_Sarsa_Prather_Denny_Becker_Hellas_Kimmel_Powell_Leinonen_2023, Šavelka_Agarwal_Bogart_Song_Sakr_2023, Wermelinger_2023}, CS2 \cite{Finnie-Ansley_Denny_Luxton-Reilly_Santos_Prather_Becker_2023, Šavelka_Agarwal_Bogart_Song_Sakr_2023}, object-oriented programming \cite{Cipriano_Alves_2023, Ouh_Gan_Gan_Wlodkowski_2023}, software engineering \cite{software_change}, and computer security \cite{Malinka_Peresíni_Firc_Hujňák_Januš_2023}. The findings from these studies demonstrate that LLMs can generate plausible solutions for a broad range of questions, although the accuracy varies. Factors influencing this accuracy include the problem's complexity and the input prompt's quality.


Several comparative studies \cite{Leinonen_Denny_MacNeil_Sarsa_Bernstein_Kim_Tran_Hellas_2023, MacNeil_Tran_Hellas_Kim_Sarsa_Denny_Bernstein_Leinonen_2023, Sarsa_Denny_Hellas_Leinonen_2022, Wermelinger_2023} have examined the code explanations produced by large language models, offering insights into their effectiveness compared to those provided by students. 
Leinonen et al. \cite{Leinonen_Denny_MacNeil_Sarsa_Bernstein_Kim_Tran_Hellas_2023} assessed OpenAI Codex's proficiency in explaining programmer-error messages and evaluating the quality of suggested code fixes, highlighting its potential for effective program debugging. Additionally, Sarsa et al. \cite{Sarsa_Denny_Hellas_Leinonen_2022}  explored Codex's utility in auto-generating programming tasks and explanations, emphasizing the need for quality control in educational contexts. 
Balse et al. \cite{Balse_Valaboju_Singhal_Warriem_Prasad_2023} explored the potential of GPT-3 to provide detailed and personalized feedback for programming assessments. The study found that GPT-3 can offer accurate feedback but occasionally provides incorrect and inconsistent responses.

In terms of user studies, Budhiraja et al. \cite{Budhiraja_Joshi_Challa_Akolekar_Kumar_2024, joshi2023interviews} conducted surveys and interviews regarding the usage and perception of ChatGPT among students and instructors in undergraduate engineering universities in India. Their study found that even though students are actively using ChatGPT to generate and debug code, brainstorm new ideas, learn new concepts, get feedback on their solutions, and create new content such as reports and emails, there are concerns about the reliability and accuracy of the responses generated by ChatGPT. Moreover, the instructors have mixed opinions regarding students being allowed to use such tools, which may hamper student learning. Lau et al \cite{Lau2023Instructor} carried out a similar study discussing the opinion of computer science instructors around the globe.

While the above-mentioned studies offer valuable insights into the broader landscape of LLM integration in education, our research distinguishes itself by taking a systematic approach to evaluate and compare four prominent LLMs: ChatGPT(3.5), Microsoft Copilot, GitHub Copilot Chat, and Google Bard. Unlike prior works that often focused on specific LLMs or limited task domains, our comprehensive approach covers a diverse range of both technical and non-technical tasks commonly encountered in undergraduate computer science education. By providing detailed insights into the capabilities and limitations of each LLM, our study contributes to a deeper understanding for students and instructors involved in computer science education for selecting the most suitable tool for various educational and professional tasks. 

Thus, our research represents a significant advancement in the discussion around the effective integration of LLMs into computer science education, building upon and extending the findings of prior works in the field.
\section{METHODOLOGY}
Our methodology involved a comprehensive evaluation of four Large Language Models (LLMs) – Google Bard \cite{Bard_announcement}, ChatGPT(3.5) \cite{chatgpt_record}, GitHub Copilot Chat\cite{Copilot_GPT_Announcement}, and Microsoft Copilot \cite{Mehdi_2023}. We consider five tasks, which are as follows:
\begin{enumerate}
    \item Code Explanation and Documentation
    \item Class Assignments
        \begin{enumerate}
            \item Programming Assignments
            \item Theoretical Assignments
            \item Humanities Assignments
        \end{enumerate}
    \item Technical Interview Preparation
    \item Learning New Concepts and Frameworks
    \item Writing Emails
\end{enumerate}
We chose these five tasks as these tasks were found to be common among undergraduate computer science students in India as discussed by Budhiraja et al \cite{Budhiraja_Joshi_Challa_Akolekar_Kumar_2024}. The dataset used in our evaluation is available on Github \cite{our_dataset}. 
Table \ref{tab:newMethodology} summarizes the high-level details about the tasks, the dataset used, and how the tasks were evaluated.

The evaluation process involved seven evaluators who were pre-final and final year undergraduate computer science students from University A and University B.\\
The \textbf{responses generated by the language models for each prompt within a given task were evaluated solely by an individual evaluator.} The analysis presented for each task is independent and does not rely on or influence the analysis of other tasks.\\
Each evaluator provided a \textbf{rating to the LLM generated responses, on a scale of 1 - 10 based on a set of pre-defined metrics, where a score of 10 signifies outstanding performance.} Furthermore, the same evaluator also conducted the qualitative analysis.

Task allocation to evaluators was based on their expertise and academic qualifications, ensuring that each evaluator had successfully completed and aced relevant coursework at their university. For instance, the task of evaluating responses for email writing was assigned to evaluators who had excelled in the Communication Course, while the task of assessing responses for technical interview preparation was allocated to evaluators with strong performance in the Algorithm Design and Analysis course. The evaluation dataset comprised prompts derived from assignments and course materials that the evaluators had previously completed as part of their coursework. This ensured the evaluators possessed the requisite domain knowledge to accurately assess the responses generated by the language models.

\subsection{\textbf{Code Explanation and Documentation}}
Computer Science undergraduates frequently undertake diverse projects encompassing classwork, research, and open-source contributions. In these tasks, encountering poorly documented code poses challenges in comprehension and modification. Leveraging Large Language Models (LLMs) proves advantageous by generating explanations for critical code segments and adding relevant code comments (wherever applicable). This approach enhances code understanding and improves the overall quality of the codebase. In this task, LLMs were prompted to explain a given code snippet by adding proper comments to the code. For this task, we prepared a dataset \cite{our_dataset_code_explanation} of 10 code samples spanning multiple programming languages, including Python, Java, C/C++, and SQL. The dataset covered essential aspects of the computer science subjects which are typically taken by students in their sophomore, junior, and senior years in India. Topics included Socket Programming, Linux Kernel Programming, Object-Oriented Programming in Java, Machine Learning in Python, and SQL queries. The LLMs were provided with the following prompt:

\noindent \textit{""" Explain the following code by adding appropriate comments to the code followed by the code snippet: <Full code snippet went here>""".} 

\begin{table*}
    \centering
    \scriptsize
    \begin{tabular}{|p{4cm}|p{4cm}|p{4cm}|p{4cm}|}
         \hline
          \textbf{Task} & \textbf{Description} & \textbf{Dataset} & \textbf{Evaluation Method} \\
         \hline
         Code Explanation \& Documentation & Add comments to code snippets or give explanation  & 10 code samples including Python, Java, C/C++, and SQL languages \cite{our_dataset_code_explanation} &  Responses rated on a scale of 1 - 10 based on clarity, placement, and reasoning in comments added to the snippets and number of attempts taken to achieve a satisfactory response. \\
         \hline
         Class Assignments & Solve programming, theoretical and humanities assignments & Assignments from Operating systems, undergraduate algorithms and Humanities class \cite{our_dataset} & Response rated on a scale of 1 - 10 based on the basis of degree of helpfulness of response (for programming and theoretical assignments) and critical analysis, informativeness and creative thinking (for humanities assignment). \\
         \hline
         Technical Interview Preparation & Generate code solutions for problems listed on LeetCode & 150 questions consisting of  diverse topics with varying levels of difficulty \cite{Neetcode} & Number of correct and incorrect solutions generated by LLMs after giving three attempts to correct the solution if the solution failed on the first try.  \\
         \hline
         
         Learning New Concepts and Frameworks & Assist in comprehending complex CS concepts and aiding in understanding frameworks and their application & 13 situational questions and interlinked concepts in CS (with interlinked subjects) \cite{our_dataset_learning_framework} & Response rated on a scale of 1 - 10 based on the depth of understanding, clarity, and coherence of response, relevance, and real-world applications and how up-to-date is the information in the response. \\
         \hline
         Writing Emails & Write emails on various scenarios faced by undergraduate computer science students & 35 email prompts based on different scenarios \cite{our_dataset} & Responses rated on a scale of 1 - 10 based on clarity, tone, relevance, and delivery of the email draft. \\
         \hline
         
    \end{tabular}
    \caption{Methodology | Scale: 1 - 10 (Higher is Better)}
    \label{tab:newMethodology}
    \vspace{-1em}
\end{table*}

The responses were rated on a scale of 1 - 10. All the parameters were given equal weightage while assigning the rating to responses.
Metrics for evaluation of the task are defined as follows:
\begin{itemize}[leftmargin=*]
\item \textbf{Clarity of comments:} How clearly the comments in the response were expressed.
\item \textbf{Appropriate placement of comments throughout the code:} Whether the comments were placed in relevant sections of the code.
\item \textbf{Logical reasoning in the comments:} The quality of reasoning and explanations provided in the comments.
\item \textbf{Number of attempts taken by the LLM to achieve a satisfactory response:} The efficiency and effectiveness of the LLM's attempts in generating a suitable response.
\end{itemize}

\subsection{\textbf{Class Assignments}}
The objective of this task was to check the proficiency of LLMs in tackling various types of assignments given to students in undergraduate computer science classes. 

\noindent\textbf{Programming Assignments.} We sourced multiple assignments from the Operating Systems class from the Monsoon 2022 semester of University A\footnote{University name not disclosed to maintain anonymity}. Operating System is a crucial part of the undergraduate computer science curriculum typically taught to sophomore or junior year students in India. It provides fundamental knowledge of hardware resources, process scheduling, memory management, file systems and understanding of other intricate low-level workings of a computer system. We took four assignments with each assignment having 2-3 questions and covering a broad spectrum of advanced topics like systems programming, inter-process communication, process scheduling, synchronization, memory management, concurrency, and kernel operations. The assignments required basic proficiency in C, assembly programming, Linux scripting, and low-level programming. Each LLM was presented with the following prompt: 

\noindent\textit{"""You are an undergraduate computer science student enrolled in operating systems course, solve the assignments and labs that will be provided next:\\
The dining philosophers problem contains five philosophers sitting on a round table can perform only one among two actions – eat and think. For eating, each of them requires two forks, one kept beside each person. Typically, allowing unrestricted access to the forks may result in a deadlock.
\begin{itemize}[leftmargin=*]
\item Write a program to simulate the philosophers using threads, and the forks using global variables. Resolve the deadlock using the following techniques: 1. Strict ordering of resource requests, and 2. Utilization of semaphores to access the resources.
\item Repeat the above system only using semaphores now with a system that also has two sauce bowls. The user would require access to one of the two sauce bowls to eat, and can access any one of them at any point of time."""
\end{itemize}
} 
\noindent Information about the deliverables required in each exercise was also provided to the LLM.
The responses were rated on a scale of 1 - 10. The metric for evaluation of the task is defined as follows:
\begin{itemize}[leftmargin=*]
\item \textbf{Degree of Helpfulness: }When evaluating the helpfulness of a response, we considered both the model rubric for the assignment (which outlines the expected criteria for the student's work) and, if the response did not align completely with the rubric, how beneficial it was in facilitating the completion of the given task. This included assessing the helpfulness of solution overviews or boilerplate generated by the LLMs. 
\end{itemize}

\noindent\textbf{Theoretical Assignments.}  This evaluation assesses the competence of LLMs in solving theoretical questions taken from the Algorithm Design and Analysis class (also called Design and Analysis of Algorithms or Undergraduate Algorithms) offered in the Winter 2022 semester at University A. The theoretical questions require the students to think and justify the proposed algorithms for a given problem. We took three assignments having 1-3 questions each on topics like Divide and Conquer, Dynamic Programming, Graphs and Network Flow.
Here is an example prompt which was provided to the LLMs: 

\noindent\textit{"""You are an undergraduate computer science student enrolled in analysis and design of algorithms course, solve the assignments that will be provided next:\\
Given an edge-weighted connected undirected graph G = (V, E) with n + 20 edges. Design an algorithm that runs in O(n)-time and outputs an edge with smallest weight contained in a cycle of G. You must give a justification why your algorithm works correctly."""}

The responses were rated on a scale of 1 - 10. The metric for evaluation of the task is defined as follows:
\begin{itemize}[leftmargin=*]
\item \textbf{Degree of Helpfulness:} Evaluations were conducted against the problem's model rubric to assess the solution's accuracy, including the correctness of the pseudocode or boilerplate provided in the response. Additionally, the rating took into account how accurately the LLM determined the time complexity of the correct solution in Big-Oh notation. In cases where the response deviated from the model rubric, the rating was assigned based on its effectiveness in helping the student reach the correct solution.
\end{itemize}

\noindent\textbf{Humanities Assignments.}  
This task was designed to test the LLMs on their capability to help students in social science and humanities subjects.  Humanities courses in computer science education promote critical thinking, ethical awareness, and effective communication, enhancing students' problem-solving abilities and fostering a holistic understanding of technology's impact on society. Regulatory bodies in Undergraduate Computer Science education in India have mandated some social science courses.

The testing dataset comprised questions from topics relating to social theorems, debate topics that are controversial yet important, ethical considerations of technology advancements and innovations, and sustainable development goals.
We assessed the LLM's ability to engage in debates on specific topics, explain theoretical concepts using practical examples, discuss Sustainable Development Goals (SDGs) with recent updates, generate creative solutions for real-world scenarios, offer balanced analyses of controversial issues, and provide insights into emerging technological advancements and innovations.
Here is an example prompt which was provided to the LLMs:\\
\noindent\textit{"""How well can you explain the Sustainable Development Goals, with regard to the current progress in the country for each goal? """}
The responses were rated on a scale of 1 - 10. All the parameters were given equal weightage while assigning the rating to responses. Metrics for evaluation of the task are defined as follows:
\begin{itemize}[leftmargin=*]
\item \textbf{Critical analysis:} This involves examining something in detail to prepare for evaluation or judgment. It explores underlying assumptions, theories, evidence, biases, and contextual factors. This metric is given a weight of 2 because it requires deeper thinking, understanding, and application of concepts beyond just presenting information, contributing significantly to the quality and depth of the response.
\item \textbf{Informativeness: } It refers to providing relevant, accurate, and valuable information. This metric is given a weight of 1 while rating a response because while it is crucial for a response to be informative, it is considered a foundational aspect and is expected in any high-quality response.
\item \textbf{Creative solutions: } These involve thinking divergently to find novel answers to problems. This metric is given the highest weight of 3 while assigning ratings for each response because creativity adds unique value and can lead to breakthrough solutions or insights, demonstrating a higher level of mastery and proficiency in problem-solving and conceptual understanding.
\end{itemize}

\subsection{\textbf{Technical Interview Preparation}}
In this task, LLMs were prompted to generate code solutions for problems listed on LeetCode.com. We selected $150$ practice questions from the roadmap provided by NeetCode\cite{Neetcode}, a popular online resource for practicing problems listed on LeetCode to help students prepare for technical interviews. Leetcode tags the problems in three difficulty levels namely easy, medium, and difficult. These questions covered diverse topics such as dynamic programming, greedy approach, divide and conquer approach, hashing, sliding window approach, stacks, linked lists, graphs, trees, etc. Here is an example prompt which was provided to each LLM: 

\noindent \textit{"""You are a computer science undergraduate student preparing for technical interviews. Please answer the below question: <Full question description went here>
"""}\\
The generated solutions were evaluated by submitting them to the Leetcode platform. If the submitted solution did not pass all the test cases on Leetcode.com, the information of the failing testcase or the error that was generated was fed back into the LLM to improve the solution. 
This iterative process continued until the solution successfully passed all the test cases or until a maximum of 3 correction attempts were made.
At the end of this process, if the solution was failing any testcase, it was marked as incorrect. Else, the solution was marked as correct.

\subsection{\textbf{Learning New Concepts and Frameworks}}

This task assesses the LLM's effectiveness in comprehending complex computer science concepts and aiding in understanding frameworks. We prepared a dataset of 13 prompts \cite{our_dataset_learning_framework}, each prompt consisting of situational questions and interlinked concepts in computer science (with interlinked subjects). Here is an example prompt which was provided to the LLM: 

\noindent \textit{"""Situation: You're on a software development team and you are tasked with learning a new framework, Django, for a web development project. You need assistance understanding the framework through explanations and guidance on reading its official documentation.\\
Question - Explanation: How effectively can you explain the key concepts and components of the Django framework, using a clear example or analogy to aid in understanding?"""}\\
The responses were rated on a scale of 1 - 10. All the parameters were given equal weightage while assigning the rating to responses. Metrics for evaluation of the task are defined as follows:
\begin{itemize}[leftmargin=*]
    \item \textbf{Depth of understanding:} How well the responses are able to comprehend computer science concepts, including intricacies and details of different frameworks.
    \item \textbf{Clarity and Coherence:} How easily interpretable were the responses and how logically did the LLM convey information, ensuring that the explanations and guidance provided were easy to follow and free from ambiguity.
    \item \textbf{Real-world Applications and Practical Relevance:} How appropriately do the LLM relate the concepts and how applicable is the information provided by LLM to real-world scenarios and practical applications in computer science.
    \item \textbf{Up-to-date Academic/Industrial Information: } The inclusion of current and relevant information from academic sources or industry standards.
\end{itemize}
\subsection{\textbf{Writing Emails}}
This task evaluated the LLMs' capability to generate content for writing emails, as email is one of the most crucial communication mediums for students in both academic and professional settings. This task evaluates LLMs on 35 prompts \cite{our_dataset_writing_emails} depicting different scenarios that an undergraduate computer science student may face at both university and industry. The prompts cover a wide spectrum of communication needs like communicating with faculty regarding coursework, feedback on projects, deadline extensions, doubts in grading, collaborating and networking with peers and personalities in academia and industry, administrative and academic queries for medical leave, scholarship, credit transfer, course and event proposals, and for personal support like forming study groups and accommodating special needs. An example prompt provided to the LLM is as follows:\\
\textit{"""As an undergraduate computer science student, draft emails on the topic that will be provided:
Write an email to the academic department that you are unwell and cannot take the upcoming exam due to a sudden illness, providing medical documentation asking for medical leave and fixing a date to accommodate your recovery
"""}
The responses were rated on a scale of 1 - 10. All the parameters were given equal weightage while assigning the rating to responses. Metrics for evaluation of the task are defined as follows:
\begin{itemize}[leftmargin=*]
    \item \textbf{Clarity:} How well the responses conveyed information in an unambiguous and interpretable manner. An example of poor and good clarity in the context of writing emails by an undergraduate student:\\Poor Clarity: 
  "I need help with the topic we learned in last lecture."\\
    Good Clarity: "I need help understanding the concept of 'Network Protocols' that we covered in the last Computer Networks lecture."
    \item \textbf{Delivery:} How effectively was the information presented. Examples of text with poor delivery and good delivery are given below.\\
    Poor Delivery: "Attached with the email is assignment. Let me know of any issues."\\
    Good Delivery: "I've attached my Assignment-2 for Data Structures and Algorithms course. If there are any issues or if you need additional information, please let me know."
    \item \textbf{Relevance:} How appropriately does the response address the details in the prompt, avoiding unnecessary information. An example is as follows:\\Poor Relevance: "I enjoyed watching the F1 Grand Prix yesterday. Also, I can't make it to tomorrow's class."\\Good Relevance: "Due to a personal commitment, I won’t be able to attend tomorrow’s ‘Machine Learning’ class. Could you please share any notes or updates that I may miss?"
    \item \textbf{Tone:} How is the style of expression in the language of response, affecting the interpersonal impact of communication. Example of a poor tone in email writing and a good tone in email writing:\\Poor Tone: "You must send me the work by noon."\\Good Tone: "Could you please send me the project report by noon?"
\end{itemize}

\section{EVALUATION}

\subsection{Quantitative Analysis}
Table \ref{tab:newEval2} shows the overall quantitative results for all the tasks. The table contains scores calculated using the average across all the ratings received for all the responses per task, along with standard deviation showcasing the fluctuation in response quality based on respective metrics of tasks. We find that different LLMs are best suited for different tasks and no single LLM performs well for all kinds of tasks. Microsoft Copilot performs the best for the task of code explanation and documentation, solving theoretical and humanities assignments. GitHub Copilot Chat performs the best for programming assignments and technical interview preparation as it is specifically trained for programming-related tasks. ChatGPT performs best for writing emails as well as generating solutions for technical intereview questions (tied with GitHub Copilot Chat). Lastly, Google Bard is the best performer for explaining new concepts and frameworks.
\\Table \ref{tab:newEval} presents the accuracy percentages of different LLMs for Technical Interview Preparation, calculated based on the number of correct solutions. In this task, Github Copilot Chat and ChatGPT achieve the highest accuracy of 96.5\% while Google Bard has the least accuracy of 57.3\%.

{\newcolumntype{C}{>{\centering\arraybackslash}p{3.8em}}
\renewcommand{\arraystretch}{1.6}

{\begin{table*}[!ht]
    \centering
    \footnotesize
    \resizebox{\textwidth}{!}{
    \begin{tabular}{|cr|CCCC|CCCC|CCCC|c|}
        \hline
        \multicolumn{2}{|c|}{\multirow{4}{*}{\textbf{LLM}}} &
            \multicolumn{4}{c|}{\textbf{Easy Questions}} &
            \multicolumn{4}{c|}{\textbf{Medium Questions}} &
            \multicolumn{4}{c|}{\textbf{Hard Questions}} & 
            \multirow{4}{*}{\parbox{5em}{\textbf{Total Accuracy \%}}}\\
            
             & & {\textbf{First Attempt Correct}} & {\textbf{Subsequent Attempt Correct}} & {\textbf{Total questions in set}} & {\textbf{Accuracy \%}} & {\textbf{First Attempt Correct}} & {\textbf{Subsequent Attempt Correct}} & {\textbf{Total questions in set}} & {\textbf{Accuracy \%}} & {\textbf{First Attempt Correct}} & {\textbf{Subsequent Attempt Correct}} & {\textbf{Total questions in set}} & {\textbf{Accuracy \%}} & \\
            \hline
            
            \multicolumn{2}{|c|}{\textbf{ChatGPT}} & 
                27 & 1 & 28 & 100 &
                87 & 3 & 93 & 96.77 & 
                19 & 1 & 22 & 90.91 & 
                 \textbf{96.5} \\
            \multicolumn{2}{|c|}{\textbf{Google Bard}} & 
                13 & 10 & 28 & 82.14 &
                30 & 26 & 93 & 60.22 & 
                0 & 3 & 22 & 13.64 & 
                 \textbf{57.34} \\
            \multicolumn{2}{|c|}{\textbf{Microsoft Copilot}} & 
                27 & 0 & 28 & 96.43 &
                68 & 1 & 93 & 74.19 & 
                13 & 1 & 22 & 63.64 & 
                 \textbf{76.92} \\
            \multicolumn{2}{|c|}{\textbf{GitHub Copilot Chat}} & 
                27 & 1 & 28 & 100 &
                87 & 2 & 93 & 95.7 & 
                19 & 2 & 22 & 95.45 & 
                 \textbf{96.5} \\
        \hline
    \end{tabular}}
    \caption{Accuracy Percentage of Large Language Models in Solving Questions on Leetcode.com}
    \label{tab:newEval}
\end{table*}}
\renewcommand{\arraystretch}{1}}

\newcolumntype{C}{>{\centering\arraybackslash}p{3.8em}}
\renewcommand{\arraystretch}{1.6}

\begin{table*}[!ht]
    \centering
    \scriptsize
    \resizebox{\textwidth}{!}{
    \begin{tabular}{|cr|CC|CC|CC|CC|c|}
        \hline
        \multicolumn{2}{|c|}{\multirow{2}{*}{\textbf{Tasks}}} &
            \multicolumn{2}{c|}{\textbf{ChatGPT}} &
            \multicolumn{2}{c|}{\textbf{Google Bard}} &
            \multicolumn{2}{c|}{\textbf{Microsoft Copilot}} & 
            \multicolumn{2}{c|}{\textbf{GitHub Copilot Chat}} & 
            \multirow{2}{*}{\parbox{10em}{\textbf{Best Performing LLM}}}\\
            
             & & {\textbf{Average}} & {\textbf{Std Dev}} & {\textbf{Average}} & {\textbf{Std Dev}}
             & {\textbf{Average}} & {\textbf{Std Dev}} & {\textbf{Average}} & {\textbf{Std Dev}} & \\
            \hline
            \multicolumn{2}{|c|}{\textbf{Code Explanation and Documentation}} & 
                8.33 & 0.52 & 7.17 & 2.56 &
                9.67 & 0.52 & 8.17 & 1.17 &
                 \textbf{Microsoft Copilot} \\
            
            \multicolumn{2}{|c|}{\textbf{Theoretical Assignments}} & 
                7.71 & 1.8 & 2.71 & 2.56 &
                8.14 & 1.07 & 6.71 & 1.89 &
                 \textbf{Microsoft Copilot} \\
            \multicolumn{2}{|c|}{\textbf{Programming Assignments}} & 
                3 & 1.93 & 5.38 & 2.39 &
                2.5 & 2.27 & 7.13 & 1.73 &
                 \textbf{Github Copilot Chat} \\
            
            \multicolumn{2}{|c|}{\textbf{Humanities Assignments}} & 
                4.83 & 3.19 & 4.5 & 3.02 &
                6.5 & 2.43 & 0.75 & 1.8 &
                 \textbf{Microsoft Copilot} \\
            \multicolumn{2}{|c|}{\textbf{Learning new concepts and frameworks}} & 
                5.33 & 3.26 & 7.75 & 2.93 &
                5.67 & 3.08 & 5 & 3.67 &
                 \textbf{Google Bard} \\
            \multicolumn{2}{|c|}{\textbf{Writing Emails}} & 
                8.29 & 1.15 & 6.06 & 1.33 &
                4.6 & 2.37 & 3.83 & 2.7 &
                 \textbf{ChatGPT} \\
            \hline
    \end{tabular}}
    \caption{Performance of Large Language Models in various tasks | Scale: 1 - 10 (Higher is better)}
    \label{tab:newEval2}
    \vspace{-3em}
\end{table*}
\renewcommand{\arraystretch}{1}

\subsection{Qualitative Analysis}

\subsubsection{\textbf{Code Explanation and Documentation}}
We find that different LLMs have different proficiency for generating code explanations and comments for different programming languages. For instance, Google Bard generates well-presented explanations with examples for SQL queries. On the other hand, it struggled with generating comments for Java and C/C++ programming languages, which was likely because of the length of code snippets involved in the evaluation. Its responses included incomplete code snippets that failed to meet the required standards, despite receiving up to five rounds of feedback from the evaluator. It performed relatively better in Python Code. Microsoft Copilot generated very detailed code explanations and comments for all the programming languages and additionally provided relevant resources from the web to search for additional information about the topic. It was also able to provide logical reasoning wherever required. For instance, it was correctly able to figure out why we can set  A + $A^T$ to 0 in the case of a skew-symmetric matrix, which Bard and ChatGPT could not. ChatGPT performed similarly to Microsoft Copilot but lagged behind in providing logical reasoning in 2 out of 10 prompts. GitHub Copilot Chat demonstrated good logical reasoning abilities in this task and was able to figure out the A + $A^T$ case as described above. However, it encountered difficulties when asked to provide in-depth explanations for code snippets, particularly when those explanations required longer and more detailed responses in plain English.

\subsubsection{\textbf{Class assignments}} In this section, we present the evaluation for each category of assignments one by one.
\\\\
\noindent\textbf{Programming Assignments.}
GitHub Copilot Chat demonstrated stronger performance among all the LLMs, showcasing its comprehensive knowledge and proficiency in the topics from the assignment set. The LLM accurately comprehended the assignment requirement, and the responses aligned with the question demand, significantly aiding the assignment completion. It showed consistency and reliability in coding-related exercises requiring good systems programming and C language knowledge. Bard's responses' were moderate in quality. The responses contained basic outlines and details that are good to be built upon by a student taking Bard’s help in assignments. At the same time, Bard diverted with the question deliverables on complex tasks like creating a custom syscall and versions of dining philosophers problem, which further damped the helpfulness of its responses. On the other hand, ChatGPT displayed a basic understanding of the tasks and interpreted the questions well, providing a high-level overview of the solution but its responses lacked details of implementation and specific Linux commands. Microsoft Copilot was the worst performer in this task. Its limitations were apparent in providing answers to exercises that required knowledge of intricate subjects such as thread scheduling, inter-process communication, syscalls, kernel modules, and versions of the dining philosophers problem. Its responses were vague and lacked requisite depth, offering limited assistance in finding the solution.
\\\\
\noindent\textbf{Theoretical Assignments.}
Microsoft Copilot demonstrated strong performance across the set of problems we used for evaluation. It was consistent in providing strong responses with correct algorithms, pseudo-code, and time complexities. GitHub Copilot Chat was competent to provide the correct pseudo-code, but it did not give the time complexity of the solution in six out of seven experiments. Despite the lack of time complexity information, its responses were found to be fairly helpful for completing the assignments. Similar challenges were found with ChatGPT as it was also giving partially correct answers.
Google Bard faced challenges in delivering correct results failing to understand the intricacies of the problem. There were instances when the output algorithm contained different procedures than those required by the question. This compromised the completeness and correctness of the solutions and gave wrong time complexities to solve the problem. For instance, a question where Bard diverted from the correct approach to solve the problem was as follows:\\ \textit{"Consider the problem of putting L-shaped tiles (L-shaped consisting of three squares) in an n × n square-board. You can assume that n is a power of 2. Suppose that one square of this board is defective and tiles cannot be put in that square. Also, two L-shaped tiles cannot intersect each other. Describe an algorithm that computes a proper tiling of the board. Justify the running time of your algorithm."}
\\The response given by Bard to the above problem describes a backtracking approach with a given time complexity of \textit{\(O\left(n^4\right)\)}, whereas the correct solution requires a Divide and Conquer paradigm based approach with the complexity of \textit{\(O\left(n^2\right)\)}.
In conclusion, the LLMs exhibited diverse patterns while tending to non-trivial questions. Microsoft Copilot delivered robust and consistently good results, while ChatGPT and GitHub Copilot Chat were vulnerable to giving only partially correct algorithms. Notably, Bard’s results were the least helpful among all the models that were evaluated.
\\\\
\noindent\textbf{Humanities Assignments.}
Across the multiple dimensions tested, ChatGPT often delivered responses with a bookish quality, lacking the critical thinking required in several questions. In contrast, Microsoft Copilot emerged as the standout performer, particularly in debate and discussion on sustainable development goals, showcasing a commendable ability to navigate and debate from multiple perspectives, as well as providing the most relevant and recent updates on current affairs. The model excelled in sociology theorems and rising innovations by delivering informative responses with critical solutions and intuitive thinking. On the other hand, Google Bard's performance varied across tasks, exhibiting strengths in real-time solutions but falling short in areas such as debate. GitHub Copilot's assistance is limited to programming-related inquiries only, as it lacked knowledge of humanities assignments, stating, "I can only assist with programming-related questions" and "My expertise is strictly limited to software development topics." As a result, it was not effective in providing assistance for humanities assignments. In summary, while ChatGPT struggled to infuse critical thinking, Microsoft Copilot emerged as the most adept in handling the complexity of such topics, offering up-to-date insights and creative solutions, thereby positioning itself as a valuable resource for students and researchers in Humanities. 

\subsubsection{\textbf{Technical Interview Preparation}}
While testing LLMs on beginner-level topics like arrays, searching, stacks, pointers, greedy and linked lists, ChatGPT, Microsoft Copilot, and Copilot Chat demonstrated commendable accuracy across difficult levels of questions, showcasing their competence in foundational concepts. However, LLMs showed varying trends when subjected to questions involving intermediate to advanced concepts like trees, tries, graphs, dynamic programming, backtracking, and heaps. Microsoft Copilot showed inconsistent performance with its responses fluctuating between being unable to answer and giving correct solutions in one go. Google Bard consistently fell short in providing accurate results in the first attempt across all the levels of topic complexity and question difficulties and even after the correction attempts of response the accuracy was still less as compared to its counterparts. GitHub Copilot Chat and ChatGPT were the best performers among all the LLMs, with high accuracy through the tests.  GitHub Copilot Chat consistently provided correct solutions across difficulty levels with soundness in solving algorithm problems, while ChatGPT showcased the most interpretable responses. ChatGPT was additionally giving easy-to-understand explanations to the solution and its approach in addition to the correct code while GitHub Copilot Chat only generated correct code with minimal explanation and comments. An example where Github Copilot Chat and ChatGPT were correct with their responses while Microsoft Copilot and Bard struggled to give correct solutions even after the third correction attempt is \textit{N-Queens Problem}, which is tagged as Hard on LeetCode website from the topic backtracking.

\subsubsection{\textbf{Learning new Concepts and Frameworks}}
Learning new concept tasks assessed how well the LLMs could help users learn new concepts by providing clear explanations and resolving any follow-up questions or doubts. ChatGPT exhibited notable strengths in certain areas, such as summarizing papers and identifying similarities and differences among them. However, it sometimes lacked critical thinking for complex questions and displayed a tendency towards a bookish style.

We also evaluated the LLM's ability to suggest programming frameworks, explain them, and provide resources for understanding them. Additionally, we assessed their performance in assisting with learning a new programming language. Google Bard demonstrated strong overall performance. For tasks that needed suggestions of references or clear and detailed information, Google Bard shined by also providing explanatory reasons. It also offers up-to-date academic/industrial information because it can access the web, which is particularly beneficial for programming frameworks that are frequently updated. The responses provided by Microsoft Copilot were clear and coherent and had up-to-date information just like Google Bard, but the answers lacked depth of understanding and were a bit too concise. In terms of up-to-date academic/industrial information, Google Bard has an advantage over Microsoft Copilot as it uses Google's superior search engine \cite{lewandowski2014evaluating} for Information Retrieval, while Copilot relies on Bing's less comprehensive search results.

In summary, ChatGPT is better at explaining new concepts in an easy-to-understand way for learning purposes. GitHub Copilot Chat struggles to give detailed explanations in plain English, specifically for non-coding tasks. However, it excels at providing responses that need technical details, such as subtle details about programming frameworks like Django. Microsoft Copilot provided current and relevant information with useful resources but lacked clarity and detail. Among the LLMs evaluated, Google's Bard demonstrated the best performance on this task. Bard consistently provided clear responses exhibiting strong logical reasoning abilities and incorporating up-to-date academic/industrial knowledge.


\subsubsection{\textbf{Writing Emails}}
Microsoft Copilot and Google Bard displayed a mix of weak and strong responses, with vagueness in responses lacking the required details as mentioned in the prompts. In nine out of thirty-five scenarios, Microsoft Copilot's responses had incomplete sentences, affecting their coherence. GitHub Copilot Chat was unable to respond to demanding prompts that involved creative composition and generating details and ideas as per the prompt. Responses were inconsistent and lacked relevance in comparison to other LLMs. This is expected as GitHub is specifically aimed at assisting in programming-related tasks. For example, when prompted with the following prompt: \textit{"""Write an email to your university's accessibility services office to request accommodations for a disability that affects your ability to complete coursework."""}, GitHub Copilot Chat denied responding to this prompt while Microsoft Copilot only gave a high-level outline of the email.
On the other hand, ChatGPT consistently delivered moderate to high-quality responses across all the prompts. It displayed good clarity, delivery, relevance and tone, effectively addressing the demand of the prompts to help undergraduate computer science students in various scenarios. ChatGPT's responses were easily distinguishable from other LLMs when the prompt required the LLM's response to be sophisticated, showcase its knowledge of the subject matter and effectively articulate it into an email. 

\section{DISCUSSION}
In our study, we conducted a comprehensive evaluation of various large language models (LLMs) frequently utilized by undergraduate computer science students, including Google Bard, ChatGPT(3.5), GitHub Copilot Chat, and Microsoft Copilot. Our findings reveal distinctive strengths across different tasks. ChatGPT and GitHub Copilot Chat exhibited remarkable proficiency in solving Leetcode.com questions across all difficulty levels. Microsoft Copilot showcased exceptional performance in providing code explanations and documentation and excelled in theoretical and humanities assignments as well. Additionally, GitHub Copilot Chat emerged as the top performer for programming assignments, while Google Bard proved to be the optimal choice for learning new concepts and frameworks. ChatGPT demonstrated the best performance in composing emails.

In this study, we anticipate that LLMs will be of significant use to students and instructors in the near future. Hence for them, it becomes increasingly important to know which LLM performs best for a given task. Our findings show that no LLM works best across all the tasks. This paper thus provides a much-needed starting point underlining the strengths and weaknesses of each LLM. We encourage students and instructors to use this study's insights along with their own assessments to make a well-informed choice best suited to their tasks and needs.

Below, we present the recommendations for students and instructors to effectively integrate LLMs into their workflow. Wherever possible, we also cite existing work where some of these recommendations have also been proposed.

\subsection{Recommendation for students:}
\begin{itemize}
    \item Content Writing: When utilizing LLMs for writing tasks such as email composition, writing cover letters, etc, it is advisable to initially create a rudimentary summary of the requirements and subsequently request the LLM to expand upon it. This methodology can help to address the potential accuracy and reliability concerns raised by \cite{Budhiraja_Joshi_Challa_Akolekar_Kumar_2024,joshi2023interviews}. They can also assist in proofreading content. Nevertheless, students are advised to consistently verify the response for any potential inaccuracies, and acknowledge their sources.
    \item Academic Tasks: LLMs can serve as an ideation and research instrument for academic tasks such as assignment completion. Instead of directly solving the assignment, they can assist students in selecting a project idea\cite{joshi2023chatgpt}, providing a fundamental outline and plan of action for the project, and comprehending problems from diverse perspectives. This could help mitigate the concerns highlighted by \cite{Lau2023Instructor}, which may adversely affect student learning outcomes.
    \item Computer Science Tasks: LLMs can be advantageous in selecting between various frameworks and technologies by presenting the pros and cons of each option\cite{software_change}. Nevertheless, students should be cognizant that LLMs may encounter difficulties with tasks necessitating logical reasoning and may provide incorrect answers as shown in our findings and \cite{Sarsa_Denny_Hellas_Leinonen_2022,Balse_Valaboju_Singhal_Warriem_Prasad_2023,Budhiraja_Joshi_Challa_Akolekar_Kumar_2024}. Instead of attempting to generate complete code, students can employ LLMs to generate code explanations\cite{Leinonen_Denny_MacNeil_Sarsa_Bernstein_Kim_Tran_Hellas_2023, MacNeil_Tran_Hellas_Kim_Sarsa_Denny_Bernstein_Leinonen_2023, Sarsa_Denny_Hellas_Leinonen_2022, Wermelinger_2023}, learn new frameworks and concepts, and suggest possible enhancements to the code. LLMs can also generate sample test cases\cite{joshi2023chatgpt}, and help to clarify any doubt about a potential solution.
\end{itemize}

\subsection{Recommendation for Instructors}
\begin{itemize}
    \item Instructors can provide guidance for using LLMs using insights shared by this study and various other guiding studies\cite{joshi2023interviews,joshi2023chatgpt,software_change,Lau2023Instructor,Budhiraja_Joshi_Challa_Akolekar_Kumar_2024,MacNeil_Tran_Hellas_Kim_Sarsa_Denny_Bernstein_Leinonen_2023}. They can try to integrate them into their courses rather than accepting the unsupervised use by students, which can negatively impact their learning\cite{software_change}.
    \item Instructors can utilize LLMs to design questions that promote critical thinking and logical reasoning skills, which are not directly solvable by LLMs as suggested by \cite{Budhiraja_Joshi_Challa_Akolekar_Kumar_2024,joshi2023chatgpt} and our results. This can help enhance students' cognitive abilities. Additionally, LLMs can be used to provide personalized questions to cater to the varying learning needs of students\cite{software_change}.
    \item Social Sciences and Humanities (SSH) assignments, are more prone to plagiarism by students as LLMs particularly excel in content and essay writing\cite{herbold2023ai}. Instructors can give specific and detailed questions that require students to cite class content and provide personal reflections on resources not commonly found on the web, such as critiques of research papers. Here, students can be encouraged to use LLMs to learn how to write critiques, read research papers, and research related topics. LLMs can then be used to provide feedback to students, enabling them to improve their work further\cite{joshi2023chatgpt,software_change}.
    \item In computer science subjects, instructors can design assignments that require students to develop real-world applications with specific requirements\cite{joshi2023chatgpt,software_change,Lau2023Instructor}, which necessitate collaboration and an in-depth understanding of multiple concepts. Students can be asked to justify their design choices and can use LLMs for ideation, brainstorming, and learning.
    \item Instructors can explore new modes of open-ended evaluations, such as giving students a topic and asking them to design a problem around it, including solutions and test cases (if applicable) \cite{joshi2023chatgpt}.
    \item Instructors and universities can collaborate with researchers working on large language models to develop LLMs specifically for the education domain. Work done by Microsoft\cite{MS_lesson_plan} has demonstrated remarkable progress in this step.
\end{itemize}

\subsection{Limitations and Future Work}
Significant advancements are happening in research related to Large Language Models and new models such as Anthropic's Claude 3, Mistral's Mixtral(8x7B), Medium and Large have been announced. The existing tools such as Google Bard\cite{Pichai_2023}, and Microsoft Copilot\cite{Gedeon_2024} are being updated with better models. However, Chen et al\cite{chatgpt_quaality_change} suggests that LLMs quality is also taking a hit with time. We plan to incorporate these advancements in our evaluation in the future.\\
One of the limitations of this study is the diversity of tasks covered, as we focused on a specific set of tasks commonly encountered by undergraduate computer science students in India. This may not fully capture the potential effectiveness of LLMs across a broader range of tasks relevant to undergraduate computer science education.\\
We were ourselves responsible for creating most of the datasets used in this study's evaluation which may have also added some undesirable bias to this study. We also acknowledge that the results and recommendations made in this paper are based on our own interpretations. We plan to evaluate these recommendations using a larger study to further establish their validity.

\section{CONCLUSION}
This study evaluated the strengths and weaknesses of ChatGPT(3.5), Microsoft Copilot, GitHub Copilot Chat, and Google Bard across various tasks relevant to undergraduate computer science education. The results indicated that different large language models (LLMs) excel at different tasks, with no single LLM outperforming the others across all tasks. These findings can guide students and instructors in selecting the most appropriate LLM for a given educational task. The study concludes by offering recommendations for students and instructors on effectively leveraging LLMs in computer science education to enhance learning outcomes.
\bibliographystyle{ACM-Reference-Format}
\bibliography{sample-base}

\end{document}